# Understanding the role and interplay of heavy hole and light hole valence bands in the thermoelectric properties of PbSe

Thomas C. Chasapis[1], Yeseul Lee[1], Euripides Hatzikraniotis[2], Konstantinos M. Paraskevopoulos[2], Hang Chi[3], Ctirad Uher[3] and Mercouri G. Kanatzidis[1*]

[1]*Department of Chemistry, Northwestern University, Evanston, Illinois, 60208, USA*

[2]*Physics Department, Aristotle University of Thessaloniki, GR- 54124, Thessaloniki, Greece*

[3]*Department of Physics, University of Michigan, Ann Arbor, Michigan 48109, USA*

**The thermoelectric properties of PbSe have significantly improved in recent years reaching figures of merit ZT~1.6. The transport properties of the hole doped high temperature thermoelectric material PbSe are particularly interesting and play a key role in this. Here they were analyzed over a wide temperature and hole concentration ranges. The special features observed in the variation of the experimental Seebeck coefficient, and Hall coefficient can be accounted for within the framework of a two-band model. Two valence bands separated by a temperature dependent energy offset are considered. The extremum of the light hole band has a density of states mass ~ 0.27$m_o$ at room temperature. It is non-parabolic and anisotropic and can be described by the Kane model. The extremum of the heavy hole band is isotropic and parabolic with a much larger density of states mass ~ 2.5$m_o$. We find that for heavily doped compositions the high mass band contributes the Seebeck coefficient even at room temperature. With rising temperature holes are transferred from the light hole to the heavy hole branch giving rise to the anomalous temperature dependent Hall coefficient which is found peaked near ~ 650 K. For Na doped samples $Pb_{1-x}Na_xSe$ for 0.01 ≤ x ≤ 0.03 the high**

**thermopower values of 200 – 300 $\mu$V/K at 900 K arise from the heavy hole band, which are responsible for the excellent thermoelectric performance of PbSe.**

## INTRODUCTION

The high thermoelectric performance (*ZT*) of a semiconductor demands high thermoelectric power to produce maximum voltage in the circuit, large electrical conductivity to minimize losses due to Joule heating and low thermal conductivity to retain maximum heat for conversion at the hot junction and avoid losses down the limbs to the cold sink [1]. The engineering of the electrical properties to achieve high power factor [2] and the reduction of the lattice component of the thermal conductivity through nanostructuring [3] are the main adopted strategies towards high thermoelectric performance in the case of lead chalcogenides.

The concept of band structure engineering stems from the complex valence band structure of Pb-chalcogenides where two inequivalent maxima, the so-called light hole band, and the lower lying heavy hole band are located at the $L$ and $\Sigma$ points of Brillouin zone respectively and the ability to control carrier access of these maxima with chemical or physical changes [4,5]. For example, the high effective hole mass of the $\Sigma$ point band enhances the Seebeck coefficient if probed by heavy doping. Band structure engineering refers to alloying on the other hand of the Pb-chalcogenide matrix to achieve an effective in tuning of the relative energies of the two bands [6-12]. Previously, extensive attention has focused on the analysis of the electrical – transport properties of $p$-type PbTe and the experimental results have been well interpreted in the framework of a two-band model; a low density of states (the light hole band) and a high density of states (the heavy hole band) which lies ~0.15 eV below in energy that contributes to the transport properties at high temperatures [5,13-18].

Lead selenide combines several attractive features relative to the tellurium analogue such as higher melting, lower cost and higher operation temperatures. The well-known PbSe system [19-23] is now attracting considerable attention as a promising thermoelectric material for high temperature power generation applications. Heavily doped *p*-type PbSe has shown *ZT*s as high as ~ 1.2 at 850 K [24,25], while after the simultaneous improvement of both electrical and thermal properties *ZT* values of ~ 1.4 – 1.6 can be obtained [9,12,25]. Theoretical calculations have suggested that hole-doped PbSe can potentially reach *ZT* ~ 2 at 1000 K [26] indicating prospects for further performance improvements. The common feature of the above studies is that high thermopower is achieved without any sign of saturation or turn-over at high temperatures due to bipolar diffusion. It is believed that the high Seebeck coefficient originates from the complex band structure and the contribution of the high density of states heavy hole band, lying deep in the valence band (much deeper than in PbTe). Under appropriate conditions of doping and temperature this heavy hole band may be exploited to enhance the thermoelectric performance of PbSe.

Based on the above, it is important to understand the electronic performance of PbSe over as broad a spectrum of carrier concentrations and as wide temperature as possible. For this reason we present here in depths study of the transport properties of heavily doped $Pb_{1-x}Na_xSe$, $0.01 \leq x \leq 0.03$. We focus on the effects of doping and temperature and we investigate the behavior of the complex valence band structure and its impact on the transport properties. Our experimental results included in our analysis cover the widest temperature range ever studied in this system. The Seebeck coefficient and the electrical conductivity were measured in the range of 300 – 900 K, while the experimental Hall coefficient covers the temperature regime of 10 – 850 K.

To our knowledge this is the first detailed application of a multi-band model for PbSe over such a broad temperature plateau. The experimental results are modelled by the adoption of two moving valence parts, a non-parabolic primary band and a parabolic secondary band, that cross each other at ~ 1500 K. As a result we can extract the partial transport properties of the two valence bands and their temperature dependence. Based on the Pisarenko plot, built with data from samples covering a wide range of doping compositions, we demonstrate for the first time that the heavy hole band plays a role even at room temperature. In particular, we show that for hole densities in the range of $2\times10^{20}$ cm$^{-3}$ the Fermi level can cross the top of the heavy hole band at room temperature. The analysis indicates that at high temperatures holes from the light hole band are transferred to the heavy hole band resulting in the anomalous temperature dependence of the Hall coefficient that it found peaked at 650 K. The heavy hole band contributes the high Seebeck coefficient values above room temperature and at 900 K essentially is the dominant contribution, accounting for the excellent thermoelectric performance of PbSe. The work described here explains previous qualitative interpretations of experimental findings and provides specific transport coefficients of the two valence bands for PbSe over the widest available temperature range.

**Transport Coefficients**

Lead chalcogenides are multivalley narrow band gap semiconductors with a direct gap at the *L* point of the Brillouin zone. Both the conduction and the valence bands are non-parabolic and anisotropic showing multiple ellipsoidal constant-energy surfaces and two components of the effective mass [4,27]: the longitudinal $m_l^*$ associated with the longitudinal axis of the ellipsoidal energy surfaces, and the transverse $m_t^*$ associated with the two transverse axes of the surfaces. The different effective masses give rise to band anisotropy defined as $K = m_l^* / m_t^*$. Due to the non-parabolicity the effective masses are functions of the energy, $\varepsilon$, defined by the interaction of electron and hole bands and the transport coefficients are described by the Kane model and the density of states as a function of energy is given by [4,27]:

$$\rho(\varepsilon) = \frac{2^{1/2} m_d^{*3/2}}{\pi^2 \hbar^3} \varepsilon^{1/2} \left(1 + \frac{\varepsilon}{E_g}\right)^{1/2} \left(1 + \frac{2\varepsilon}{E_g}\right) \tag{1}$$

where $m_d^* = N^{2/3} (m_l^* m_t^{*2})^{1/3}$ is the total density of states mass, $N$ is the number of the equivalent or degenerated valleys and $E_g$ is the $L$ point fundamental gap. The relaxation time is inversely proportional to the density of states. Also the mean square matrix element of the interaction of a carrier with a scatterer decreases (due to the non-parabolicity) with increasing energy in the case of acoustical and optical phonon scattering [4,27].

The mechanisms of charge carrier scattering in Pb-chalcogenides are both temperature and concentration dependent. For low temperatures, the scattering by the Coulomb and the short range potential of vacancies are expected to dominate, while around or above room temperatures the polarization scattering by optical phonons, the collision between carriers, the deformation potential of acoustic phonons and the deformation potential of optical phonons are among the premier scattering mechanisms. Additionally, as a result of the complex band structure, the scattering of carriers between different valleys or bands is also expected to contribute [27-31]. Despite the complexity and the number of scattering mechanisms, the high temperature transport properties of heavily doped semiconductors may be described assuming the deformation potential of acoustic phonons as the dominant scattering mechanism, which is the case in the present study. The deformation potential of acoustic phonons with that of optical phonons and the intervalley scattering actually produce an effective deformation potential coefficient [27,31]. For lead chalcogenides the expression of the acoustic phonon relaxation time in the framework of the Kane model after taking into account the energy dependent interaction matrix is written [4,29,30,32]:

$$\tau_{ac}(\varepsilon) = \tau_0\left(\varepsilon + \varepsilon^2\alpha\right)^{-1/2}\left(1 + 2\alpha\varepsilon\right)^{-1}\left[1 - \frac{8\alpha\left(\varepsilon + \varepsilon^2\alpha\right)}{3\left(1 + 2\varepsilon\alpha\right)^2}\right]^{-1} \quad (2)$$

where $\varepsilon = E / k_B T$ is the reduced energy and $\alpha = k_B T / E_g$ is the band non-parabolicity parameter. At equilibrium, the carrier distribution follows the Fermi-Dirac statistics, and is a function of the reduced energy and the reduced chemical potential $\eta$ [27]:

$$f(\varepsilon) = \frac{1}{1 + \exp(\varepsilon - \eta)} \quad (3)$$

By defining the generalized Fermi integrals as [31,33]:

$$^{n}F_{k}^{m}(\eta) = \int_0^\infty \left(-\frac{\partial f}{\partial \varepsilon}\right)\varepsilon^n(\varepsilon + \varepsilon^2\alpha)^m\left[(1 + 2\varepsilon\alpha)^2 + 2\right]^{k/2} d\varepsilon \quad (4)$$

the transport coefficients are expressed as follows:

the Seebeck coefficient:

$$S = \frac{k_B}{e}\left[\frac{^{1}F_{-2}^{1}}{^{0}F_{-2}^{1}} - \eta\right] \quad (5)$$

the carrier density, which is proportional to the effective density of states $N_C$:

$$p = \mathrm{N}_C \, {}^0F_0^{3/2} = \frac{(2m_d^* k_B T)^{3/2}}{3\pi^2 \hbar^3} \, {}^0F_0^{3/2} \tag{6}$$

the mobility:

$$\mu = \frac{e\langle \tau_{ac}(\varepsilon) \rangle}{m_c^*} = \frac{2\pi\hbar^4 e C_l}{m_c^* (2m_b^* k_B T)^{3/2} E_d^2} \frac{3\,{}^0F_{-2}^1}{{}^0F_0^{3/2}} \tag{7}$$

where $k_B$ is the Boltzmann constant, $\hbar$ is the reduced Planck's constants and $e$ is the electronic charge. $C_l$ is the parameter determined by the combination of the elastic constants and $E_d$ is the deformation potential [4,31].

The Hall factor:

$$A = \frac{3K(K+2)}{(2K+1)^2} \frac{{}^0F_{-4}^{1/2} \cdot {}^0F_0^{3/2}}{\left({}^0F_{-2}^1\right)^2} \tag{8}$$

The Hall coefficient $R_H$ is related to the carrier density $p$ and the Hall factor $A$ though the equation:

$$R_H = A/ep \tag{9}$$

The masses $m_b^*$ and $m_c^*$ entered into the mobility expression are the density of states mass and the conductivity mass of a single ellipsoid respectively and both are related to the total density of states mass through [34]:

$$m_b^* = N^{-2/3} m_d^* \tag{10}$$

and

$$m_c^* = m_d^* N^{-2/3} \frac{3K^{2/3}}{2K+1} \tag{11}$$

where $N$ is the number of the equivalent valleys and $K$ is the band anisotropy defined previously.

**Main electronic properties of PbSe**

Previous reports on electrical, optical and thermoelectric properties of both *n*- and *p*- type PbSe focused mainly on low temperatures and carrier densities and have shown similarities between the properties of *n*- and *p*- type PbSe [19-24,33,35-43]. The properties however deviate substantially at higher carrier concentrations and temperatures.

Transport and optical investigations on *n*-type PbSe have confirmed the non-parabolic dispersion law of the conduction band [33,35,37,40,44]. The number of equivalent valleys is $N=4$ and the band anisotropy parameter $K$ lies in the range of 1.75 – 2 [36,37,45]. A room temperature value of $m_d^* \sim 0.27m_o$ was found to fit the concentration dependent Seebeck coefficient (the Pisarenko plot) of *n*-type PbSe assuming a single band Kane model and its temperature dependence follows the power law $m_d^* \propto T^{0.5}$ [33]. The temperature dependence of Hall mobility of *n*-type single crystals revealed two slopes; up to ~ 500 K a slope of ~ – 2 was observed indicating band non parabolicity, while for higher temperatures an increased slope of ~ – 3 was associated with possible optical phonons excitations [37]. The value of 25 eV was proposed for the effective deformation potential coefficient at room temperature after the analysis of the temperature dependent transport properties of *n*-type PbSe within the framework of the Kane model and the assumption of acoustic phonon scattering [33].

As with *n*-type samples, the room temperature Pisarenko plot of *p*-type PbSe was fitted by the incorporation of a single band model and the room temperature effective mass for the $L$ point maximum was found $m_{dL}^* \sim 0.27 - 0.28m_o$. Similar effective masses were obtained by the assumption of either single non parabolic [19] or single parabolic band model [24]. However, discrepancies were observed concerning the carrier density regime at which the agreement between the theoretical and the experimental behavior was achieved [19,24]. The analysis of the temperature dependent thermopower in the framework of a single band Kane model has shown that up to ~ 450 K the power law of $m_{dL}^{*} \propto T^{0.5-0.8}$ was found to hold, while for higher temperatures a strong rise of the effective mass was observed [19]. Two slopes were also observed in the temperature dependence of the Hall mobility; up to ~ 500 K the mobility scales with the temperature following $\mu \propto T^{-(1.5-2.8)}$, while for higher temperatures it scales according to $\mu \propto T^{-(4.2-4.6)}$ [19]. Pei *et* al. assumed that a higher deformation potential coefficient, i.e. 35 eV, of the *L* point valence band might be responsible for the lower room temperature mobility of the *p*-type compositions relative to the respective *n*-type ones [2]. Additionally, the comparison between *n*- and *p*- type PbSe with carrier densities in the range of $1\times10^{20}$ cm$^{-3}$ has shown that the Seebeck coefficient is comparable up to ~ 450 K, with the *n*-type showing considerably lower values at higher temperatures [24].

The above studies indicate that at relatively low carrier densities and temperatures the behavior of the transport properties of *p*-type PbSe may be explained satisfactorily by a single band model resembling those of *n*-type analogue [2,33]. The significant deviations from the single band behavior observed at high carrier densities and temperatures actually demonstrate the necessity for the presence of an additional extremum located deeper in the valence band. The presence of the second hole band is supported by the temperature dependent Hall coefficient suggesting the conduction of two types of carriers [21]. At sufficiently high doping and temperature this second extremum dominates the transport properties of *p*-type PbSe.

A simplified depiction of the band structure of PbSe is shown in Fig. 1 (a). Two symmetric bands at the $L$ point of the Brillouin zone separated by a direct energy gap $E_g$, and a second valence band located at the $\Sigma$ point with a valence band offset $\Delta E$ are assumed. The $L$ point valence band with degeneracy $N_{VL}$ = 4 is denoted as the light hole (lh) band and the highly degenerate $\Sigma$ point valence band with $N_{V\Sigma}$ = 12 [2,22] as the heavy hole band (hh). Theoretical calculations based on the generalized gradient approximation (GGA) have shown a flat, high mass band at ~ 0. 35 eV (0 K) below the band edge [26] while the quasiparticle self-consistent approach (QSGW) point to a secondary band maximum between the $\Gamma$ and $X$ points of the Brillouin zone [46]. The room temperature direct gap is ~ 0.26 eV [47,48] and the valence band offset is valued ~ 0.25 eV [21] . As it is shown in Fig. 1 (b) on raising temperature the direct gap increases with a rate of ~ $10^{-4}$ eV/K [9,49] and the light hole band moves downwards resulting in the decrease of the valence band offset and in band convergence at ~ 1500 K [21]. Recent spectroscopic studies (Angle Resolved Photoemission Spectroscopy, ARPES) of the electronic band structure of PbSe revealed an unusual temperature dependent relative movement between the valence band maxima and a band convergence at ~ 1300 K [50]. The convergence of the these valence bands and the increase in the effective valley degeneracy at higher temperatures through the appearance of the heavy hole band on the Fermi surface is believed to be the reason for the deviation from the single band model behavior reported above, and for the unique thermoelectric properties of *p*-type PbSe.

In order to extract the ″hidden″ transport coefficients of the heavy hole band and their temperature dependence we analyzed the experimental transport properties of heavily doped *p*-type PbSe in the framework of the two band model as illustrated in Fig. 1 (a). The material parameters used for modelling the experimental data is summarized in Table I. For the temperature dependent $L$ point direct gap (see Fig. 1 (b)) we used the experimental values obtained by diffuse reflectance spectroscopy measurements on PbSe [9,48]. The relative position of the two inequivalent valence band maxima of Figs. 1 (a) and (b) was obtained from the analysis of the temperature dependent absorption coefficient of doped PbSe after the subtraction of the free carrier contribution [9,21]. The effective mass of the light hole band was assumed to follow the same temperature dependence obtained for *n*-type PbSe [33]. The band degeneracy values of 4 and 12 were assumed for the light and heavy hole bands respectively [2,22]. Similarly to *n*-type PbSe, the light hole band is assumed non-parabolic and anisotropic with a temperature independent anisotropy factor $K$ of 1.75 [33,36,37] where the heavy hole band is taken, for simplicity, as parabolic and isotropic. The elastic constants were taken to be temperature independent and equal to 9 GPa [31].

**The Two-Band Model**

In the presence of two bands the total transport coefficients are determined by the transport coefficients of the individual components. The transport coefficients of the light hole band were calculated from Eqs. (4) – (11). The corresponding coefficients of the heavy hole band were obtained by the respective equations of the $L$ band accounting for the parabolic dispersion law by setting $a = 0$ in Eq. (4), and writing the reduced chemical potential as $\eta_\Sigma = \eta_L - \Delta E/k_B T$ , where $\Delta E$ is the valence band offset defined in Fig. 1 (a).

For two band semiconductors the total Seebeck and Hall coefficients from both bands is given [13,51]:

$$S = \frac{S_L \sigma_L + S_\Sigma \sigma_\Sigma}{\sigma_L + \sigma_\Sigma} = \frac{S_L f + S_\Sigma}{f + 1} \qquad (12)$$

$$R_H = \frac{R_{HL}\sigma_L^2 + R_{H\Sigma}\sigma_\Sigma^2}{(\sigma_L + \sigma_\Sigma)^2} = \frac{R_{HL} f^2 + R_{H\Sigma}}{(f+1)^2} \tag{13}$$

where $\sigma_L$ , $\sigma_\Sigma$ are the partial electrical conductivities, $S_L$ , $S_\Sigma$ are the partial Seebeck coefficients and $R_{HL}$, $R_{H\Sigma}$ are the partial Hall coefficients. The parameter *f* accounting for the conductivity ratio is given by [13]:

$$f = f_0 \frac{{}^0F^1_{-2}(\eta_L, \alpha)}{{}^0F^1_{-2}(\eta_\Sigma, 0)} \tag{14}$$

$$\sigma_{total} = \sigma_L + \sigma_\Sigma = \sigma_\Sigma (f+1) \tag{15}$$

where the constant $f_0$ is proportional to the ratio of the relaxation times and the effective masses of the two bands [13].

For the analysis of the room temperature Pisarenko plot the above equations are simplified by using the relation $\sigma_i = e p_i \mu_i$ and defining a constant mobility ratio $b = \mu_L / \mu_\Sigma$

$$S = \frac{S_L p_L b + S_\Sigma p_\Sigma}{p_L b + p_\Sigma} \tag{16}$$

$$R_H = \frac{A_L p_L b^2 + A_\Sigma p_\Sigma}{e(p_L b + p_\Sigma)^2} \tag{17}$$

where $p_L$, $p_\Sigma$ and $A_L$, $A_\Sigma$ are the partial hole densities and Hall factors of the *L* and *Σ* bands respectively [4,15,52].

## EXPERIMENTAL

Ingots with nominal compositions $Pb_{1-x}Na_xSe$ ($0 \leq x \leq 0.04$) were prepared by mixing appropriate ratios of reagents into carbon-coated quartz tubes. The tubes were sealed under vacuum (~$10^{-4}$ Torr), heated to 1150 °C over a period of 12 h, soaked at that temperature for 5 h, slowly cooled to 1080 °C over 12 h, and rapidly cooled to room temperature over 3 h. Compositions with $0.005 \leq x \leq 0.03$ were further processed. They were ground by a mechanical mortar and pestle to reduce the particle size < 53 $\mu m^3$ and densified at 650°C for 10 min under uniaxial press of 60 MPa in Argon atmosphere. The samples were cut and polished into a parallelepiped with dimensions of ~ 2 mm × 3 mm × 10 mm. The electrical conductivity and Seebeck coefficient were measured simultaneously under a helium atmosphere (~ 0.1 atm) from room temperature to 900 K using an ULVAC-RIKO ZEM-3 system. Room temperature Hall coefficients were measured in a home-built system in a magnetic field ranging from 0.5 to 1.25 T, utilizing simple four-contact Hall-bar geometry, in both negative and positive polarity of the magnetic field to eliminate Joule resistive errors. For the heavily doped SPS compositions, x = 0.01, 0.02 and 0.03, temperature dependent Hall measurements were carried out in a Quantum Design MPMS system (10 - 300 K) and/or an air-bore Oxford superconducting magnet (300 - 850 K) equipped with a Linear Research ac bridge (LR - 700). The room temperature transport coefficients of the studied compositions are tabulated in Table II.

## RESULTS AND DISCUSSION

Before proceeding to the application of the two band model, we address that the assumption of a Kane – type light hole band is actually based on the room temperature concentration dependence of the Hall mobility (see Table II) [2,19,33]. We found that the consideration of a non – parabolic light hole band using Eqs. (6) – (9) (not shown) predicts better the experimental results compared to a parabolic one [33]. The necessity for a lower lying secondary heavy hole band was confirmed by the application of the single non – parabolic band model to the temperature dependent transport properties using Eqs. (5) and (9). The obtained temperature and concentration dependences of the effective mass in conjunction with the anomalous temperature dependence of the experimental Hall coefficient [15,18,19,21,53] indicated that a two band rather than a single band model is the most appropriate. Our results within the framework of a single band model were found in good agreement with already published ones [2,19,24,33].

### Room temperature Pisarenko plot and hole density distribution in the two valence bands

In order to determine the effective mass of the heavy hole band we analyzed the data on the room temperature Pisarenko plot, see Fig. 3 (a). Table II summarizes the Hall density and the Seebeck coefficient of the $Pb_{1-x}Na_xSe$ compositions used to construct the $S$ versus $p_h$ plot. As can be seen from Table II the x values cover a wide range, $0 \le x \le 0.04$, and the final compositions yield hole densities in the range ~ $2\times10^{18}$ – $6\times10^{20}$ $cm^{-3}$. Both polycrystalline ingots and SPS processed samples were used. Differences are observed between ingots and SPS samples concerning the hole density for the same amount of Na doping.

In Fig. 2 we compare the measured carrier concentrations together with the theoretical values obtained by the assumption that each substitutional Na atom creates one hole. It is clear from Fig. 2 that the best agreement between experimental and calculated hole densities is found for the SPS samples. The systematic deviation of the ingots' values is attributed to the incomplete solubility as a result of partial Na segregation [54]. Micro-FTIR studies of several polycrystalline ingots (not shown in this work) revealed large variations of the free carriers plasma frequency for the same amount of Na. It has been shown that Na-rich precipitates formed in PbTe during cooling to room temperature are re-dissolved in the matrix during heating [55]. Since the SPS samples are processed at 650 $^{o}$C we assume that re-dissolution of the precipitates occurs giving higher hole densities relative to the ingots for the same amount of Na.

Figure 3 (a) shows the experimental Seebeck coefficients as a function of the Hall carrier density at room temperature. As can be seen, the measured Seebeck coefficient shows dual behavior; (a) it drops from ~ 230 $\mu$V/K for the lightly doped compositions to ~ 30 $\mu$V/K as the hole density reaches the value of ~ $1\times10^{20}$ cm$^{-3}$ and (b) saturates at around 26 – 29 $\mu$V/K when the carrier concentration is > $1\times10^{20}$ cm$^{-3}$.

This situation is similar to the one observed in *p*-type PbTe [4,14,56,57]. For *p*-type PbTe the flattening of the Pisarenko plot at ~ 50 $\mu$V/K for hole densities greater than ~ $5\times10^{19}$ cm$^{-3}$ was explained by a two-band model of a light and a heavy hole band. It was established that at low hole densities the light mass band dominates the thermopower, while at high hole densities as the Fermi level moves deeper into the valence band, the heavy band contribution is responsible for the saturation of the Seebeck coefficient. This was found to be true for both K-doped [39] and Na-doped PbTe. Despite these similarities, the PbSe case is not completely settled.

Although both theoretical and experimental studies confirm the complex nature of the valence band structure of PbSe [21,50,58] the picture concerning the contribution of the heavy hole band at room temperature is still ambiguous. Wang *et* al. based on the Pisarenko plot of Na-doped PbSe [9,24] suggested that the contribution is negligible up to hole densities ~ $3\times10^{20}$ cm$^{-3}$ and a single band model was found to explain well their experimental data. Zhang *et* al [39] proposed that in the case of K-doped PbSe up to ~ $1.5\times10^{20}$ cm$^{-3}$ the main contribution comes from the light hole band, while Vinogradova *et* al.[19] stressed the deviation from a single band model for hole densities higher than $2\times10^{20}$ cm$^{-3}$.

In this work we studied compositions with very high hole densities reaching the values of ~ $6\times10^{20}$ cm$^{-3}$, (see Table II). Since the saturation of the thermopower for hole densities ~ $2\times10^{20}$ cm$^{-3}$ is an experimental fact we conclude that for the heavily doped *p*-type PbSe compositions the heavy hole band does contribute to the transport even at room temperature. Using Eqs. (16) and (17) we performed the analysis and calculated the room temperature effective masses of the two bands, the partial Seebeck coefficients and the partial hole densities as a function of hole doping.

The parameters of the *L* band were determined by employing the single Kane model and Eq. (5), and adjusting the density of states mass to obtain the best fit to the experimental values up to Hall densities ~ $1\times10^{20}$ cm$^{-3}$. The Seebeck coefficient, shown as blue line in Fig. 3 (a), was calculated using an effective mass $m_{dL}^{*}$= 0.27m$_0$ for the light hole band. This value of $m_d^*$ is in good agreement with the one proposed by Wang *et* al. [33] and Vinogradova *et* al. [19] after the Pisarenko analysis of Na-doped PbSe samples considering a single band model and acoustic phonon scattering. An effective mass of 0.28m$_o$ was found to fit the Pisarenko relation in the low hole density limit in the case of K-doped PbSe [39].

In Fig. 3 (a) we also plot the concentration dependent Seebeck coefficient for several *n*-type PbSe compositions. Considering the $PbCl_2$-doped materials [37] the experimental values are in fairly good agreement with the theoretical ones, suggesting that the effective mass, $m_{dL}^{*}$ = $0.27m_o$, is the same for the conduction and the light hole valence bands of PbSe. A similar conclusion was reached by Wang *et* al. after the comparison of Br-doped PbSe with the *p*-type analogues up to $1\times10^{20}$ $cm^{-3}$ [33]. Accordingly, Chernik *et* al.[59] suggested that the conduction and the valence bands of PbSe have similar parameters up hole densities of $\sim 1\times10^{20}$ $cm^{-3}$. The last column of Table II tabulates the chemical potential values obtained by the fitting of the Pisarenko plot with the single Kane model. As can be seen, the chemical potential increases with Na amount and moves from the *L* point band edge for the undoped PbSe to ~ 0.33 eV deeper in the valence band when the hole density reaches $\sim 6\times10^{20}$ $cm^{-3}$. With the density of states mass of the *L* band at $0.27m_0$, there are three adjustable parameters that define the total Seebeck coefficient (Eqs. (16) and (17)) namely; the energy difference between the two branches, $\Delta E$, the heavy hole band effective mass and the mobility ratio of the two sub-bands.

Starting with the valence band offset $\Delta E$ the room temperature value of ~ 0.25 eV was proposed based on the absorption coefficient spectra of hole doped PbSe (see Table I) [21,22]. In order to confirm this value we followed the analysis performed by Kolomoets *et* al. for *p*-type PbTe [13]. Assuming a single band model we calculated the density of states mass for each of our compositions using the Eqs. (5) and (6). We observe that up to $\sim 1 - 2\times10^{20}$ $cm^{-3}$ the obtained effective mass is almost constant ~ 0.23 – 0.27 $m_0$, while for higher hole densities there is an abrupt increase. This increase of the effective mass actually denotes the failure of the single band model due to the contribution of the high mass heavy hole band. According to Kolomoets *et* al. the value of the chemical potential at the knee point of the dependence of the effective mass on carrier density is actually the room temperature valence band offset [13]. As can be seen from Table II the chemical potential is ~ 0.21 – 0.25 eV for hole densities in the range $1 - 2\times10^{20}$ $cm^{-3}$ (at the knee point of the dependence of the effective mass on carrier density) quiet close to the valence band offset predicted by the respective Equation of Table I. At the same range of hole densities the Seebeck coefficient saturates (see Fig. 3 (a)), meaning that the chemical potential at the onset of the flattening of the Pisarenko line is essentially the band offset in the case of a two band model.

Concerning the density of states mass of the heavy hole band, different values have been proposed in the literature. Based on optical investigations of *p*-type PbSe, Veis *et* al. proposed the value ~ $2.5m_o$ [22] while the same value was found to describe the flattening of the Pisarenko plot of the $K_xPb_{1-x}Se$ at high hole densities [39]. Gurieva *et* al. have adopted a value of ~ $2m_o$ to describe the temperature dependence of the Hall coefficient of *p*-type PbSe [17]. Theoretical calculations of the band structure of PbSe yielded an average mass ~ $0.48m_0$ for one ellipsoid [46], which based on Eq. (10) and assuming a band degeneracy of 12 gives for the total effective mass the value of ~ $2.5m_0$. On the other hand, Veis *et* al. [21], based on the temperature dependent Hall coefficient, and Wang *et* al. [9], after modelling the high temperature transport coefficients of $Pb_{1-x}Sr_xSe$ on the assumption of a mixed scattering mechanism, suggested a room temperature value of ~ $4.5m_o$.

The determination of the heavy hole band effective mass from the best fitted Seebeck data in the concentration range where the Pisarenko plot flattens (Fig. 3 (a)), was found to be more challenging. This is because the total Seebeck is co-determined by the heavy hole effective mass and the mobility ratio, in a way that the best fit yields higher effective mass for higher mobility ratio. However, as we shall show later, the temperature dependent transport properties are well described by adopting the value of ~ $2.5m_o$ for the heavy hole band, in accordance with Veis *et* al.

[22] and Zhang *et* al. [39]. Consequently, the best fit to the experimental data using the total Seebeck coefficient i.e. the black line in Figure 2(a), was obtained by adjusting the room temperature mobility ratio, *b*, to the value of ~ 4. The dashed red line of Fig. 3(a) shows the concentration dependent thermopower of the heavy hole band as obtained from the two band model. The Seebeck coefficient of the $\Sigma$ band is higher than that of the *L* band over the whole concentration range, owing to the higher effective mass. As expected, the total Seebeck is a weighted average of the individual bands and lies closer to the Seebeck values of the *L* band.

Based on the fitting of the Pisarenko plot, Fig. 3 (b) displays the room temperature relative population of the individual bands as obtained by the partial hole densities. As can be seen, the fraction of holes in the light hole band is almost unity up to a critical concentration of ~ $1\times10^{20}$ $cm^{-3}$. Up to this concentration range the Seebeck coefficient of Fig. 3(a) matches a Pisarenko relation calculated from the parameters of the light hole band alone (the blue line of Fig. 3(a)), while for higher hole densities the thermopower starts to saturate. Based on these observations, the critical *L* band population may be assumed to be ~ $1\times10^{20}$ $cm^{-3}$ holes, since up to these hole densities the $\Sigma$ band may be considered to be almost empty. For higher hole densities the chemical potential (see Table II) moves deeper in the valence band and the heavy hole band starts to acquire a steady fraction of holes. For ~ $2\times10^{20}$ $cm^{-3}$ holes the chemical potential crosses the band edge of the $\Sigma$ band, $\eta = \Delta E$, and for ~ $6\times10^{20}$ $cm^{-3}$ holes the chemical potential is located below the heavy hole band edge, $\eta$ = 0.33 eV, (Table II). These findings actually demonstrate why the heavy doping is necessary to probe the high density of states heavy hole band in PbSe.

The room temperature saturation thermopower for *p*-type PbTe was found around 50 $\mu$V/K and the Pisarenko analysis based on a two band model suggested only ~ $5\times10^{19}$ $cm^{-3}$ holes are sufficient for the Fermi level to reach the $\Sigma$ band maximum [14,17]. This difference in the light hole band population between PbTe and PbSe arises from the lower $L$ /$\Sigma$ energy offset in PbTe [17].

**Partial transport coefficients**

After the determination of the room temperature parameters, we extended the application of the two band model to higher temperatures in order to describe the temperature dependence of the transport coefficients of the three heavily doped $Pb_{1-x}Na_xSe$ compositions with x = 0.01, 0.02 and 0.03. The measured temperature dependent Seebeck coefficients, electrical conductivities and Hall coefficients are displayed in Fig. 4. The experimental Seebeck coefficients show typical behavior of degenerate semiconductors and increase with the temperature (Fig. 4(a)). All the samples show comparable room temperature values ~ 25 $\mu$V/K while for higher temperatures the thermopower scales with the Na amount. For the x = 0.01 composition the Seebeck coefficient reaches ~ 200 $\mu$V/K at 900 K, while the x = 0.02 and x = 0.03 compositions show ~ 170 and ~ 160 $\mu$V/K respectively. The room temperature electrical conductivities are high ~ 2900 – 3100 S/cm. The conductivity decreases with increasing temperature and has higher values with higher Na amounts over most of the temperature range (Fig. 4 (b)). The temperature dependent Hall coefficients, $R_H$, (Fig. 4 (c)) show typical two band behavior [4,15,53]. Up to ~ 300 K the Hall coefficient is almost constant; for higher temperatures it increases reaching a maximum at ~ 650 K and then decreases. The experimental values scale with the Na content in the whole temperature range, i.e. higher Na amount yields lower Hall coefficient. Using the low temperature values we obtained the apparent hole densities ($p = 1/R_{H10K}$) for the x = 0.01, 0.02 and 0.03 samples, i.e. ~ $1.8\times10^{20}$ cm$^{-3}$, $4.7\times10^{20}$ cm$^{-3}$ and $7.6\times10^{20}$ cm$^{-3}$ respectively.

The application of the two-band model requires knowledge of the temperature dependences of all parameters involved in the calculations. Those are summarized in Table I. The fitting was performed to the experimental transport coefficients of Fig. 4 using the Eqs. (12) – (15). The reduced chemical potential and the conductivity ratio $f_0$ were adjusted to fit the temperature dependent Seebeck and Hall coefficients. The studied compositions are extrinsic heavily doped semiconductors with no sign of bipolar contribution. Thus, the analysis assumes that at each temperature the neutrality condition holds in the form [13,18,57]:

$$p = p_L + p_\Sigma = const. \tag{18}$$

where $p$ is the apparent hole density obtained by the low temperature Hall coefficient ($p$=1/$R_{H10K}$).

The results of our analysis are displayed in Figs. 5 – 7. For each composition the temperature dependent chemical potential (Fermi levels, $\eta$) and their relative position within that valence band structure are summarized in Figs. 5(a) and (b) respectively. The partial transport coefficients of the two valence bands are included as red and blue lines – symbols and the best fit total Seebeck and Hall coefficients are denoted as black solid lines in Figs. 6 and 7.

We start our discussion with the best fit obtained chemical potential values of Fig. 5 (a). In this way we can illustrate the effect of doping and the temperature on the extracted transport coefficients of the two valence bands. The room temperature chemical potentials ($\eta$ values) of the studied compositions are in the range of ~ 0.2 – 0.35 eV and in good agreement with the respective values listed in Table II. The small discrepancies observed are attributed to the fact that the $\eta$ values of Table II are obtained using the room temperature Hall coefficient, while those of Fig. 5 (a) are obtained using the Hall coefficient at ~ 10 K with the ratio $R_{H300K}$ / $R_{H10K}$ ~ 1.12. On raising the temperature to 900 K the chemical potential decreases. For the x = 0.01 composition $\eta$ is almost zero at 900 K, while for the x = 0.02 and 0.03 samples ~ 0.15 eV. The two most heavily doped compositions show a sharper decrease of the chemical potential in the whole temperature range relative to the x = 0.01 sample. For the latter a slightly different slope is observed for temperatures higher than ~ 500 K.

In Fig. 5 (b) we present the relative position of the chemical potentials obtained after subtracting the $\eta$ values from the line that represents the position of the light hole band. It is clear that for the x = 0.01 sample the chemical potential lies higher in energy than the heavy hole band maximum in the whole temperature range. However, the difference between the two levels is in the range of ~ 1.6 – 1.2 $k_B T$ in going from 300 to 900 K. For the x = 0.02 and 0.03 samples the chemical potential is lower than the heavy hole band at room temperature. At ~ 600 K the Fermi level is crossing the heavy hole band for the x = 0.02 composition and for the x = 0.03 the crossing is observed at ~ 900 K.

Figures 6 (a) – (c) display the temperature dependence of the partial hole densities for the x = 0.01, 0.02 and 0.03 samples. As can be seen, with increasing temperature there is a monotonic reduction of the light hole band carrier density following the decrease of the chemical potential. This gives rise to the corresponding increase of the heavy hole band carrier concentration as a result of the neutrality condition. The rate of the carrier's density variation is proportional to the valence band offset reduction, i.e. $10^{-4}$ $cm^{-3}$/K. For the x = 0.01 composition (Fig. 6 (a)) the room temperature hole densities for the two valence bands are ~ $1.6\times10^{20}$ $cm^{-3}$ and ~ $1\times10^{19}$ $cm^{-3}$ and the light hole band is expected to contribute most. At higher temperatures holes are excited to the heavy hole band so at ~ 800 K both bands show the same number of holes ~ $8\times10^{19}$ $cm^{-3}$. For the x = 0.02 the hole densities are ~ $3\times10^{20}$ and $1.5\times10^{20}$ $cm^{-3}$ for the light and heavy hole band respectively. The equality in hole population is now observed at ~ 500 K (Fig. 6 (b)). The room temperature chemical potential value for the x = 0.03 sample (Fig. 5 (a)) is found ~ 0.32 eV and both of the bands show comparable population of ~ 3 – $4\times10^{20}$ $cm^{-3}$ (Fig. 6 (c)). For this composition the crossing of the hole densities is at ~ 400 K and the heavy hole band carrier density at 900 K is ~ $5.5\times10^{20}$ $cm^{-3}$. The major points from Figs. 5 and 6 (a) – (c) are: i) the position of the chemical potential is lowered as a result of hole doping in the whole temperature range, ii) at high temperatures the chemical potential moves to the band gap and holes are transferred form the light to heavy hole band and iii) the crossing point where holes exist in both bands is concentration dependent and shifts from ~ 800 K to ~ 400 K when the total hole density increases from ~ $2\times10^{20}$ (x = 0.01) to ~ $8\times10^{20}$ $cm^{-3}$ (x = 0.03) respectively.

Figures 6 (d) – (f) display the partial Seebeck coefficients and the calculated thermopower for the best two band model fit for each sample composition. As can be seen, the agreement between the model and the experimental values is good in the whole temperature range. Considering first the x = 0.01 composition of Fig. 6 (d) the experimental Seebeck follows very close the calculated Seebeck coefficient of the light hole band up to ~ 500 K. The slope of the experimental Seebeck coefficient changes slightly at ~ 600 K. For higher temperatures the experimental thermopower is larger than that of the light hole band suggesting that the $\Sigma$ band contribution is restricted mainly to temperatures higher than ~ 500 K. On raising the temperature to 900 K the $L$ band thermopower changes from ~ 30 to 160 $\mu$V/K while that of the $\Sigma$ band decreases from ~ 350 $\mu$V/K and saturates at ~ 300 $\mu$V/K at high temperatures. The decrease of the reduced chemical potential (Fig. 5 (a)) gives rise to the increase of the light hole band thermopower with rising temperature. The saturation of the heavy hole band Seebeck is the result of the variation of the respective chemical potential $\eta_\Sigma$. The latter is found negative in the whole temperature range and is nearly saturated for temperatures higher than ~ 500 K.

For the two most heavily doped compositions x = 0.02 and 0.03 the situation is different (Figs. 6 (e) and (f)) relative to the x = 0.01 sample discussed above. Here the change of the slope of the experimental *S vs T* plot is clearly more pronounced. The Seebeck coefficient of the *L* band is lower than the measured ones in the whole temperature range. For these samples the contribution of the heavy hole band is pronounced even at room temperature, where the chemical potential is lower than the heavy hole band edge (Fig. 5 (a)). This explains the Pisarenko plot of Fig. 3 (a) where Seebeck flattening for hole densities beyond ~ $2\times10^{20}$ $cm^{-3}$ is observed. For both of these samples the carrier density of the heavy hole band at room temperature is higher than $1\times 10^{20}$ $cm^{-3}$ (Figs. 6 (b) and (c)). For the x = 0.02 composition the Seebeck coefficient of the light hole band varies from ~ 10 – 120 $\mu$V/K while for the x = 0.03 composition from ~ 80 –100 $\mu$V/K. In each case the heavy hole band displays increasing Seebeck values with rising temperature, suggesting a nearly degenerate band. For the x = 0.02 sample the thermopower varies from ~ 150 – 230 $\mu$V/K while for the x = 0.03 changes from ~ 100 – 200 $\mu$V/K.

Our results on the partial Seebeck coefficients clearly demonstrate the effect of the complex valence band structure on the thermopower of PbSe. The contribution of the high mass – high Seebeck secondary band produces significantly higher Seebeck values of the *p*- type compositions relative to the *n*-type analogues discussed above. For doping levels up to ~ $1.5\times10^{20}$ $cm^{-3}$ holes the heavy hole band is found under-populated, considering its high density of states, and shows nearly constant Seebeck values ~ 300 $\mu$V/K above 600 K. Under heavy doping conditions, > $4\times10^{20}$ $cm^{-3}$ holes, the behavior of the heavy hole band resembles that of a nearly degenerate one displaying increasing Seebeck values with temperature. Due to the increased effective density of states the heavy hole band reaches ~ 200 $\mu$V/K for a population ~ $5\times10^{20}$ $cm^{-3}$ holes at 900 K, when the light hole band shows ~100 $\mu$V/K for ~ $2\times10^{20}$ $cm^{-3}$ holes (Fig. 6 (e)). In general, the total Seebeck coefficient is expected to lie between the partial Seebeck coefficients of the two bands. Figures 6 (d) – (e) demonstrates that depending on doping level the thermopower values as high as ~ 250 $\mu$V/K are expected at 1000 K. Additionally, it is now clear that decreasing the valence band offset one can decrease the temperature in which the heavy hole band begins to contribute [9,12].

The comparison between the calculated and the experimental Hall coefficients is shown in Figs. 7 (a) – (c). In each case the agreement is found to be good. At this point we need to stress that a temperature independent heavy hole mass of $2.5m_0$ was found adequate to fit all the transport coefficients for the x = 0.01 and x = 0.02 compositions. For the x = 0.03 sample this assumption yielded a poor fit to the experimental Hall coefficient (the maximum at ~ 650 K could not be reproduced). For this reason it was necessary to assume a temperature dependent heavy hole effective mass, i.e. $d \ln m^* / dT = 0.3$. One reason for this dependence could be the complex nature and anisotropy of the heavy hole band [46].

The redistribution of carriers between the two valence bands give rise to an anomalous temperature dependence of the Hall coefficients (Figs. 7 (a) – (c)) [5,18,21,38]. As we have already shown, holes are transferred form the light to heavy hole band as a result of shifting of the chemical potential with raising temperature. Up to ~ 650 K the increase of the Hall coefficient is related to the decrease of the light hole density and to the high mass – low mobility heavy holes which contribute to a small amount [4]. Figure 7 (d) presents the energy independent term of the conductivity ratio, $f_0$, (see Eq. (14)) for the three studied compositions. As can be seen, $f_0$, decreases with rising temperature and scales with the amount of Na up to ~ 500 K. At ~ 650 K (maximum of the Hall coefficient) $f_0$ is almost unity for the three compositions. A further increase in temperature makes $f_0$ lower than unity, decreasing continuously and converge irrespective of the hole density.

## CONCLUDING REMARKS

We have presented a comprehensive study of the transport properties of *p*-type PbSe thermoelectric material which shows that the complex valence band structure of the material has profound impact on the thermoelectric performance. The effects of the heavy doping, and temperature, on both the Seebeck coefficient and the Hall coefficient were determined in detail. We have shown that the experimental transport coefficients may be well described with two moving inequivalent band maxima; a light hole band and a lower laying heavy hole band. The room temperature Pisarenko plot was described by adopting a non-parabolic light hole band with effective mass ~ 0.27 $m_0$ that contributes most up to ~ $1\times10^{20}$ $cm^{-3}$ holes, and a parabolic heavy hole band with an effective mass of ~ 2.5 $m_0$ responsible for the flattening of the Pisarenko plot for hole densities higher than ~ $2\times10^{20}$ $cm^{-3}$. The valence band offset determined by the chemical potential at the saturation point of the Pisarenko plot was found ~ 0.25 eV. On raising temperature the effective mass of the Kane – type light hole band increases resembling the behavior of the *n*-type analogue. The temperature dependence valence band offset demonstrated that the crossing of the two bands takes place at high temperatures ~ 1500 K. The relative movement of the two bands resulted in the transferring of holes from the light hole band to the heavy hole band giving rise to an anomalous temperature dependent Hall coefficient peaked at ~ 650 K. The assumption of a parabolic heavy hole band with a temperature independent effective mass seemed to hold up to ~ $4.7\times10^{20}$ $cm^{-3}$ holes. For hole densities higher than ~ $7\times10^{20}$ $cm^{-3}$ such an assumption yielded a poor fit and a temperature dependent effective mass was found to describe better the experimental results. This might be an indication of the possible complexity and anisotropy of the heavy hole band. Our analysis yielded the partial transport coefficients of the two valence bands over the widest temperature range of 300 – 900 K. At 900 K the heavy hole band can reach Seebeck values as high as 200 – 300 $\mu$V/K accounting for the high thermopower of *p*-type PbSe, and the outstanding thermoelectric performance of the material. We strongly believe that the extracted temperature and concentration dependent partial transport coefficients may serve as a guide for future band structure engineering towards the design of optimized PbSe-based thermoelectric materials. Our work demonstrates the general similarity of the temperature dependence of the valence band structures of PbSe and PbTe. However, the larger valence band offset of the selenium analogue accounts for the significant distinction and the different performance of the two lead chalcogenides thermoelectrics.

**Acknowledgments**

This work was supported as part of the Revolutionary Materials for Solid State Energy Conversion, an Energy Frontier Research Center funded by the U.S. Department of Energy, Office of Science, Office of Basic Energy Sciences, under Award No. ED-SC 0001054.

******Corresponding Author*: m-kanatzidis@northwestern.edu

## Figure Captions

**FIG. 1:** (a) A schematic representation of the complex band structure of PbSe. Two bands at the *L* point of the Brillouin zone (blue curves); the conduction band and the light hole band (lh), separated by a direct energy gap $E_g$. The heavy hole band (hh) at the $\Sigma$ point of the Brillouin zone with a valence band offset $\Delta E$ is denoted by the red curve, (b) The relative positions of the bands as a function of temperature. The lines are obtained by the respective equations of Table I that describe the variation of $E_g$ and $\Delta E$ with temperature.

**FIG. 2**: Comparison of the room temperature hole density ($p_h = 1/R_H$) of the ingots and SPS processed $Pb_{1-x}Na_xSe$ compositions against Na content. The dashed line indicates the theoretical hole density obtained by the assumption that each Na atom creates one hole.

**FIG. 3:** (a) Room temperature Pisarenko plot. The Seebeck values of both *p*- and *n*- type PbSe are displayed. The black solid line corresponds to the two-band model calculated Seebeck coefficient. Dashed lines represent the concentration dependent Seebeck coefficients of the light hole band (blue line) and the heavy hole band (red line), (b) The relative population of the two bands as a function of the total hole density $p_{total} = p_{lh} + p_{hh}$.

**FIG. 4:** Temperature dependent Seebeck coefficient (a), electrical conductivity (b) and Hall coefficient (c) of the $Pb_{1-x}Na_xSe$ compositions with x = 0.01, 0.02 and 0.03. Note the different scale of the $R_H$ axis between the x = 0.01 and x = 0.02 and x = 0.03 materials.

**FIG. 5:** (a) The Fermi level as a function of temperature obtained by the application of the two-band model for the x = 0.01, 0.02 and 0.03 compositions, (b) the Fermi level relative to the position of the light hole and the heavy hole valence bands.

**FIG. 6:** (a) – (c) The temperature dependent hole densities of the two valence bands for the $Pb_{1-x}Na_xSe$ compositions. Blue symbols indicate the light hole band and red symbols indicate the heavy hole band, (d) – (f) Comparison between the experimental thermopower (open circles) and the two-band model calculated Seebeck coefficient (black solid line) of the x = 0.01, 0.02 and 0.03 materials. The obtained partial Seebeck coefficients of the light hole band (blue solid lines) and the heavy hole band (red solid lines) are also shown.

**FIG. 7:** Comparison between the experimental (open symbols) and the calculated (black solid lines) Hall coefficient of the x = 0.01 (a), x = 0.02 (b) and x = 0.03 (c) $Pb_{1-x}Na_xSe$ compositions, (d) The energy independent conductivity ratio $f_0$ as a function of temperature for the three studied materials. Note that for T ~ 650 K the conductivity ratio convergences to unity.

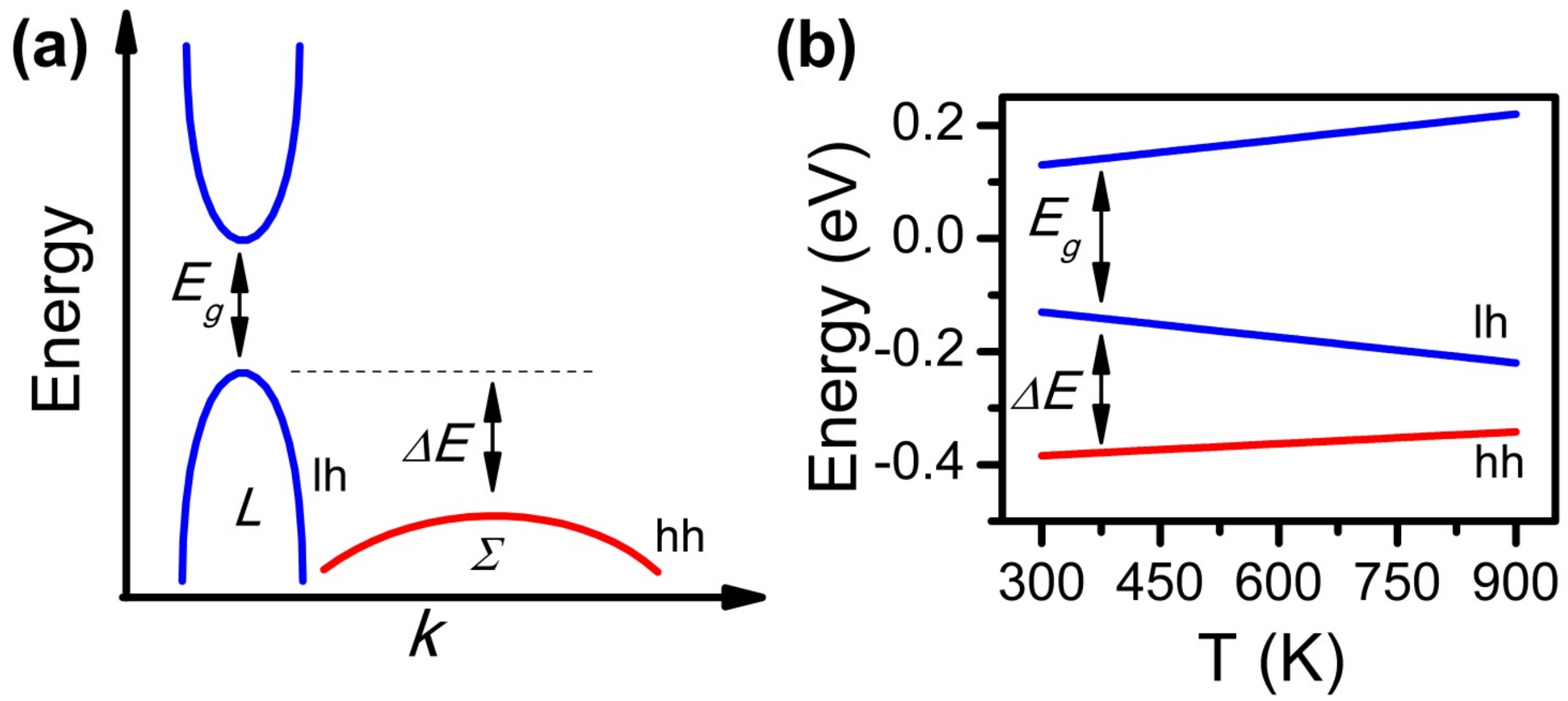

**Figure 1**

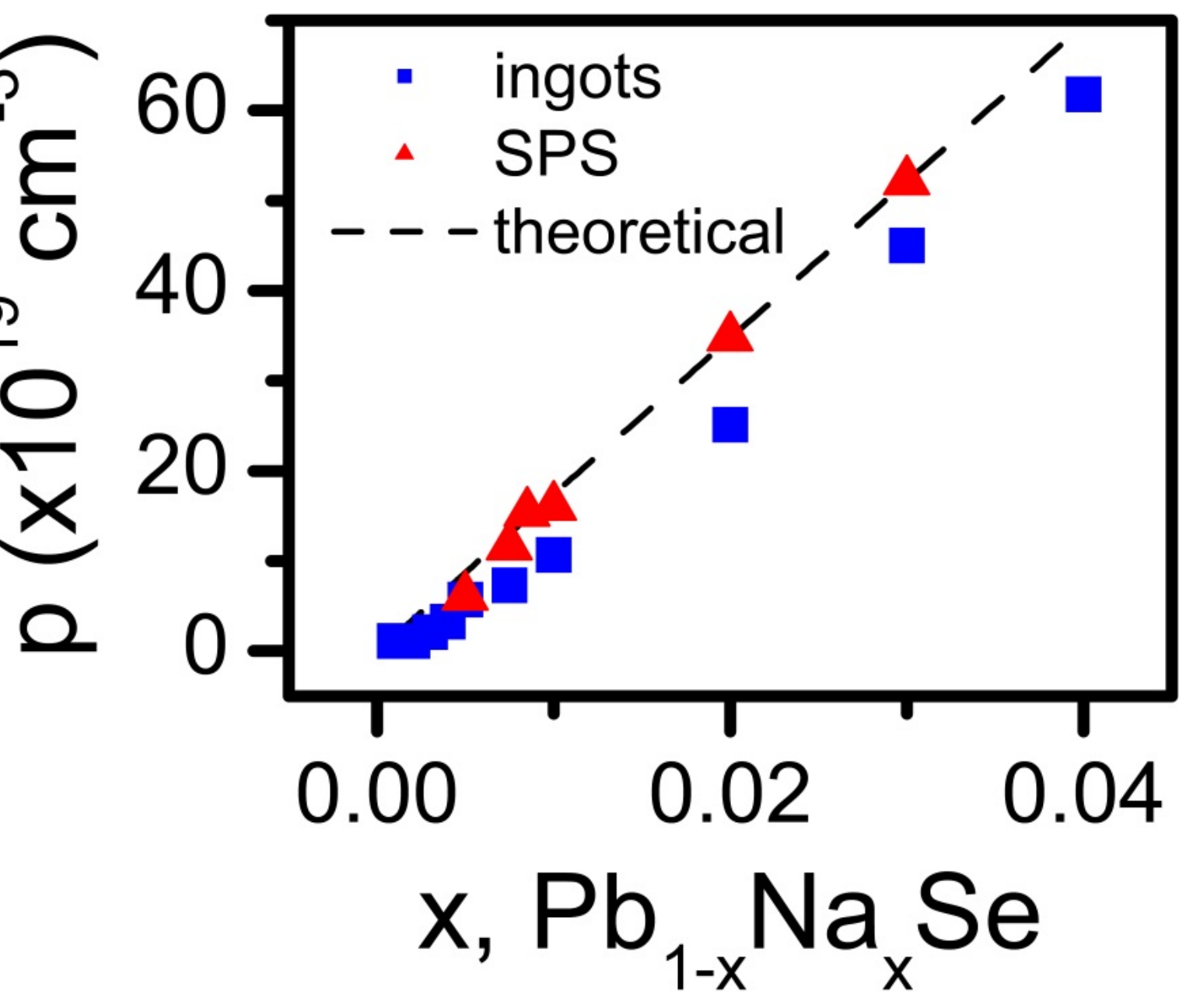

**Figure 2**

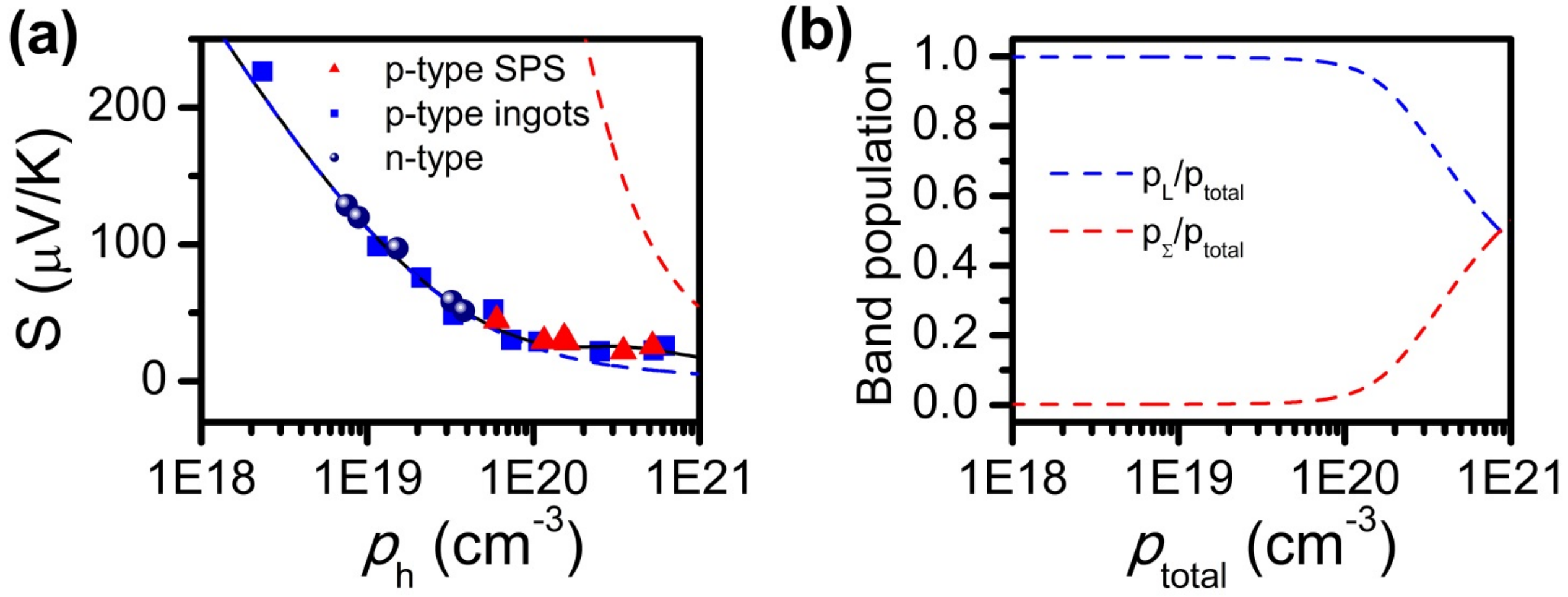

**Figure 3**

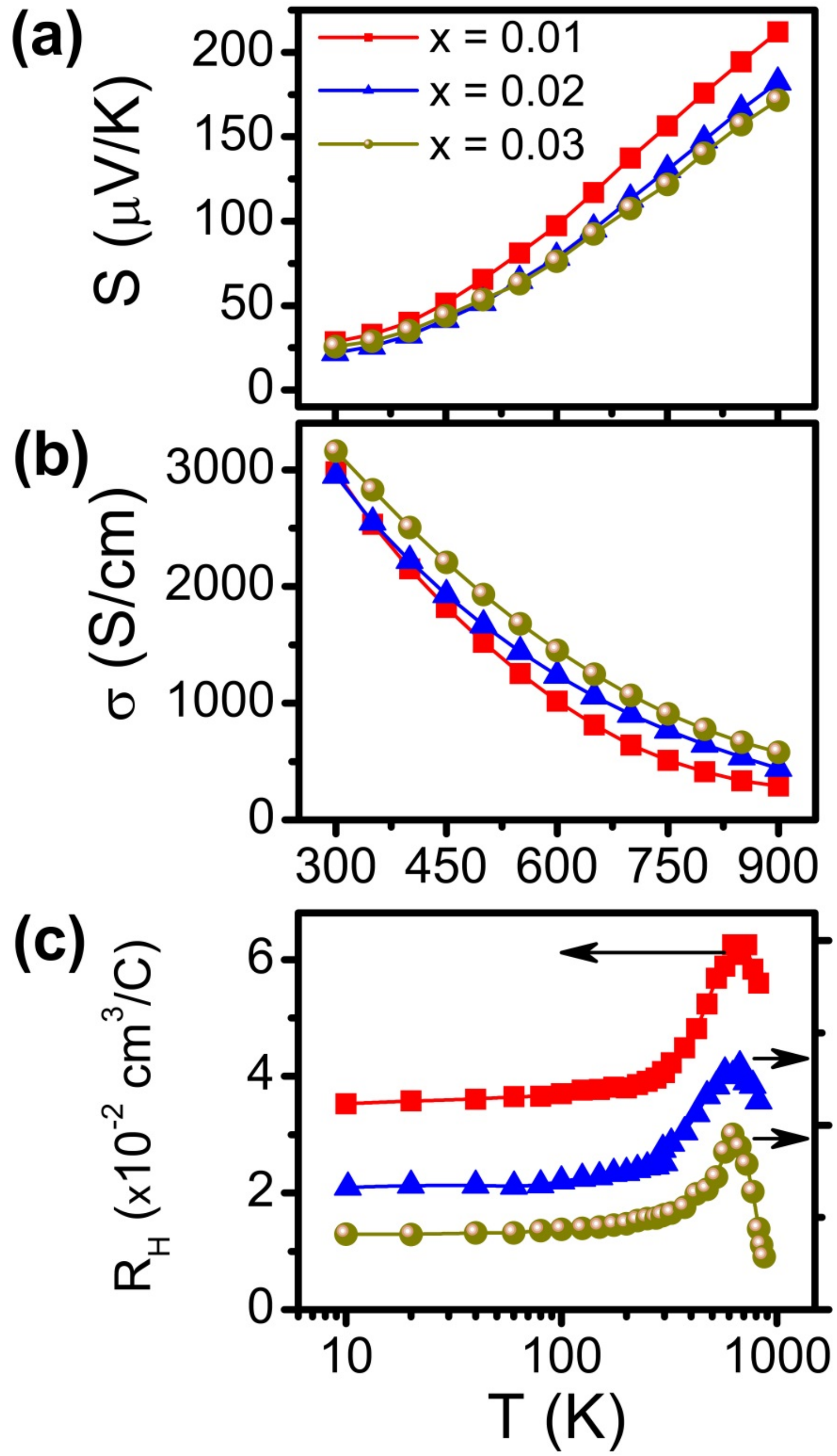

**Figure 4**

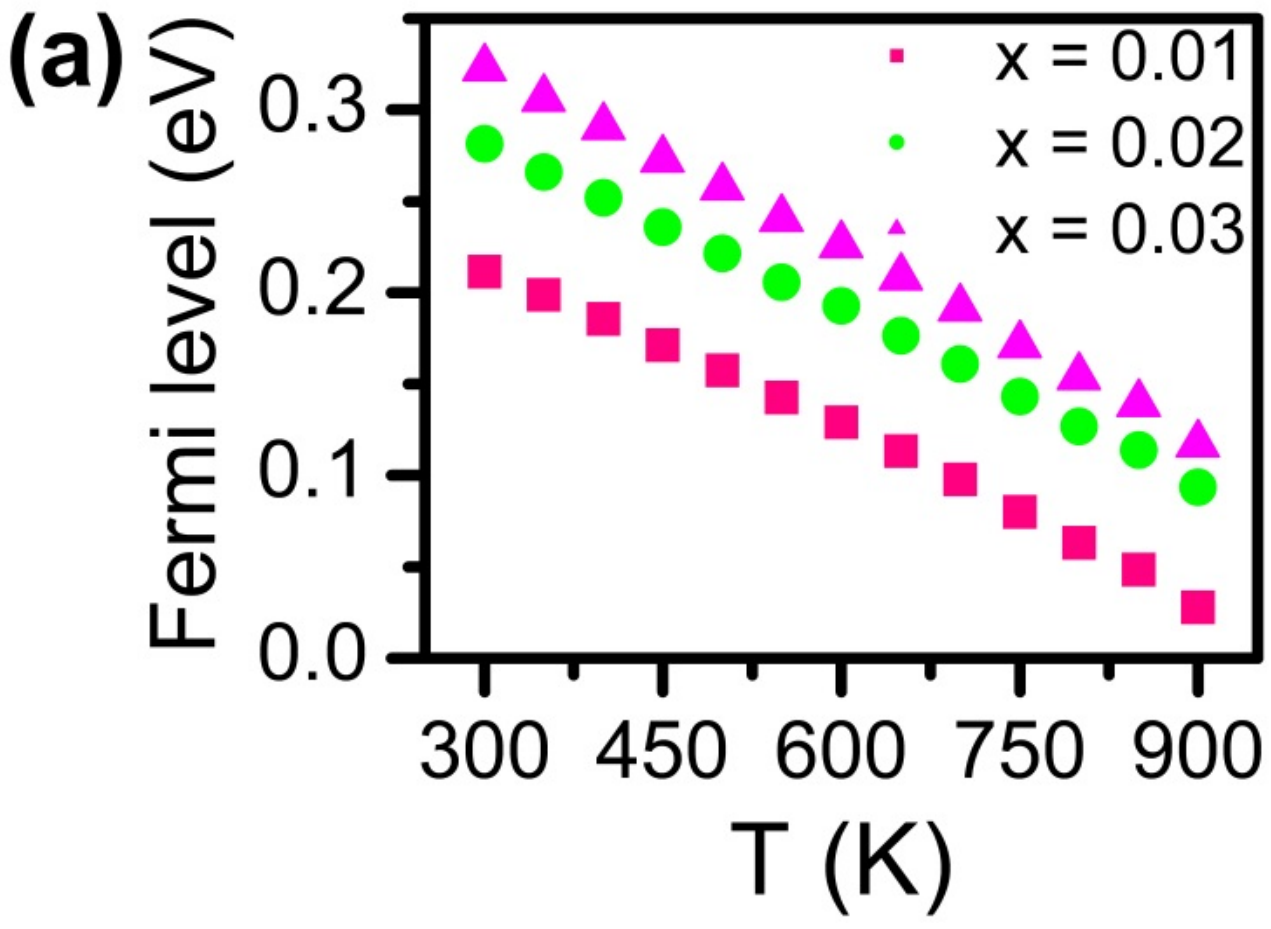

**Figure 5**

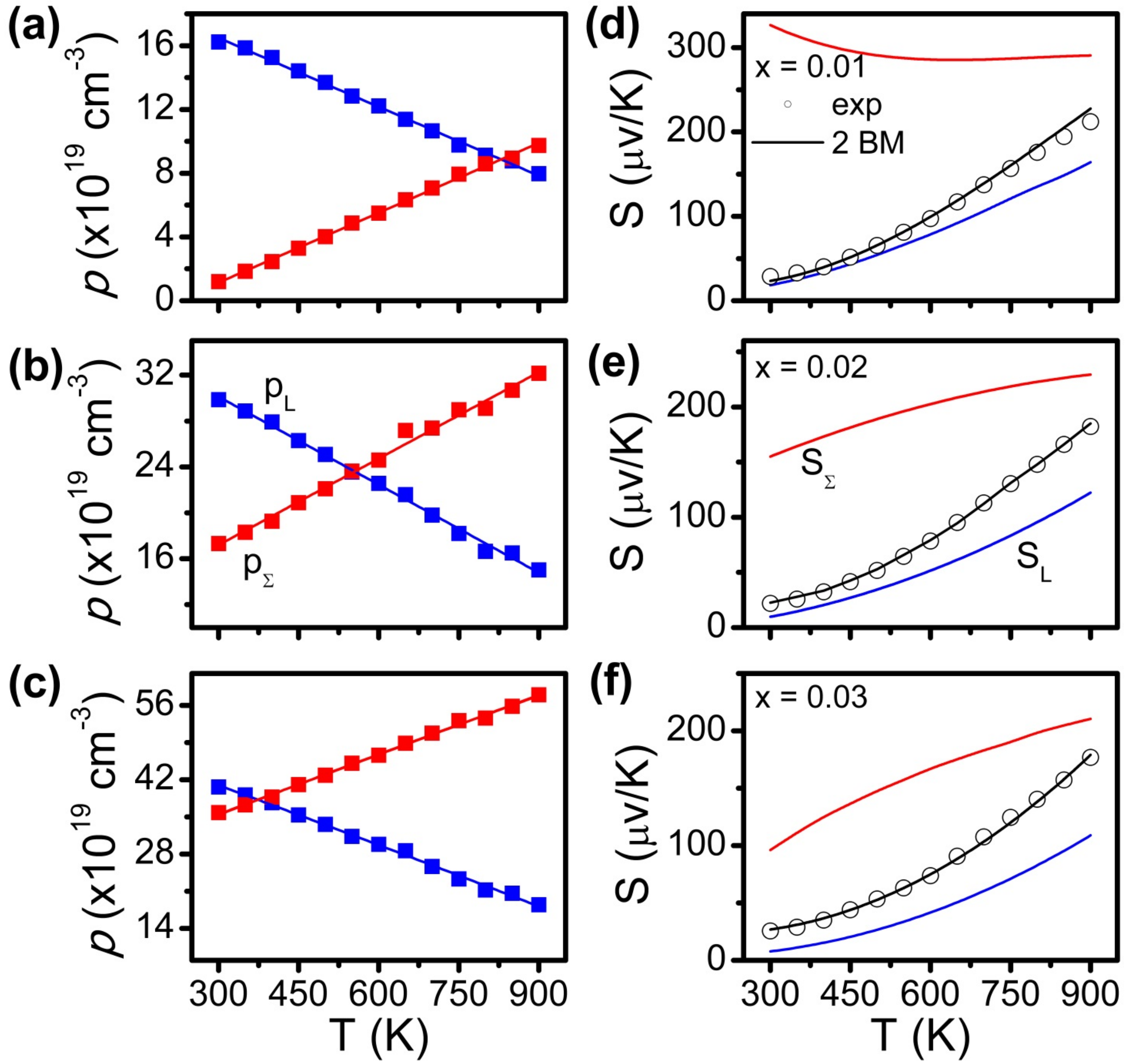

**Figure 6**

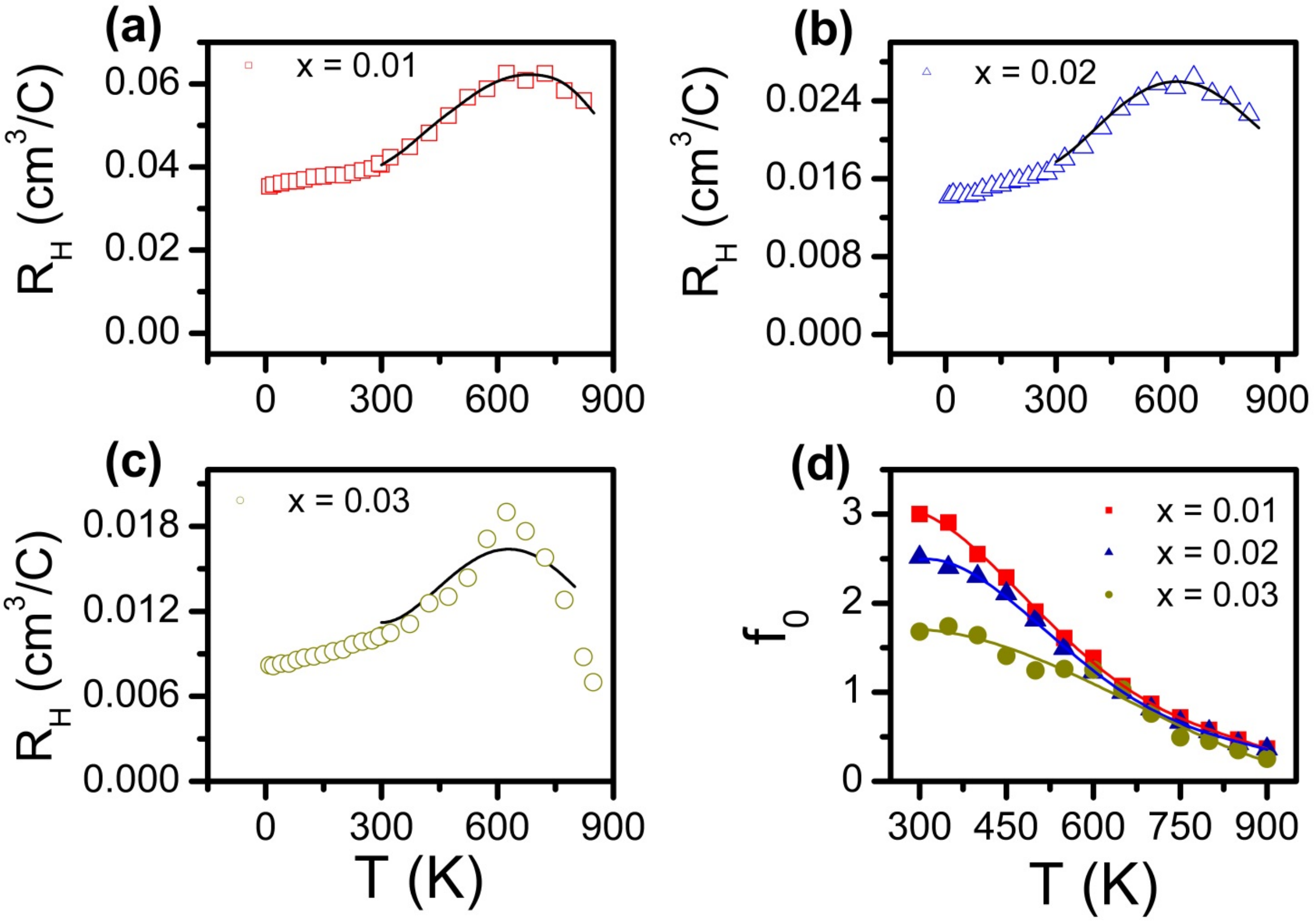

**Figure 7**

**Table I:** Material parameters used for modeling the temperature dependent transport properties

| Parameter | Value | Reference |
|---|---|---|
| *L*- point Energy gap, $E_g$ (eV) [a] | $0.17 + 3.0\cdot10^{-4}T$ | [9,48] |
| Valence band offset, $\Delta E$ (eV) | $0.32 - 2.2\cdot10^{-4}T$ | [9,21] |
| Light hole (lh) effective mass $m_{lh}$ ($m_0$) [b] | $0.18 + 3.0\cdot10^{-4}T$ | [33] |
| Heavy hole (hh) effective mass $m_{hh}$ ($m_0$) | ~ 2.5 | this work |
| *L*-, *Σ*-band degeneracies | 4, 12 | [2,33] |
| Lattice constant (Å) [c] | 6.12 | this work |
| Elastic constant $C_l$ (N/m$^2$) | $9\cdot10^9$ | [31] |

[a] we use the temperature dependent gap of the undoped PbSe; the analysis is performed within the rigid body approximation

[b] a linear fit to the m *vs* T dependence of Ref. [33] was considered

[c] obtained after the refinement of the powder X-ray diffraction pattern

**Table II:** Room temperature hole density ($p_h = 1/R_H$), Seebeck coefficient ($S$), Hall mobility ($\mu_H$) and electrical conductivity ($\sigma$) of $Pb_{1-x}Na_xSe$ compositions. The last column is the chemical potential ($\eta$) obtained after fitting the room temperature Pisarenko plot.

| **Composition** | | $p_h$ **(×10^19 cm^-3)** | $S$ **(μV/K)** | $\mu_H$ **(cm²/Vs)** | $\sigma$ **(S/cm)** | $\eta$ **(eV)** |
|---|---|---|---|---|---|---|
| PbSe | (ingot) | 0.23 | 226.3 | 1295 | 483 | -0.01 |
| $Pb_{0.999}Na_{0.001}Se$ | (-//-) | 1.15 | 98.8 | 986 | 1854 | 0.05 |
| $Pb_{0.997}Na_{0.003}Se$ | (-//-) | 2.11 | 75.7 | 765 | 2583 | 0.07 |
| $Pb_{0.996}Na_{0.004}Se$ | (-//-) | 3.29 | 48.4 | 553 | 2912 | 0.10 |
| $Pb_{0.995}Na_{0.005}Se$ | (ingot/SPS) | 5.74 / 6.02 | 52.5 / 44.5 | 314 / 224 | 2889 / 2155 | 0.13 / 0.14 |
| $Pb_{0.9925}Na_{0.0075}Se$ | (-//-) | 7.33 / 11.6 | 30.3 / 29.5 | 247 / 146 | 2903 / 2720 | 0.15 / 0.19 |
| $Pb_{0.9915}Na_{0.0085}Se$ | (SPS) | 15.3 | 31.7 | 116 | 2840 | 0.21 |
| $Pb_{0.99}Na_{0.01}Se$ | (ingot/SPS) | 10.7 / 16 | 28.9 / 28.6 | 170 / 116 | 2922 / 2981 | 0.18 / 0.21 |
| $Pb_{0.98}Na_{0.02}Se$ | (-//-) | 25.1 / 34.8 | 21.8 / 21.9 | 82 / 51 | 3264 / 2861 | 0.25 / 0.28 |
| $Pb_{0.97}Na_{0.03}Se$ | (-//-) | 45.0 / 52.1 | 22.4 / 25.6 | 52 / 38 | 3726 / 3159 | 0.31 / 0.32 |
| $Pb_{0.96}Na_{0.04}Se$ | (ingot) | 61.8 | 26.07 | 26 | 2606 | 0.33 |